\documentclass[sigconf,nonacm]{acmart}
\AtBeginDocument{%
  }

\usepackage{amsmath,amssymb,amsfonts}
\usepackage{graphicx}
\usepackage{subcaption}
\usepackage{textcomp}
\usepackage{xcolor}
\usepackage{multirow}
\usepackage{algorithm}
\usepackage{algpseudocode}
\usepackage{pgfplots}
\usepackage{amsmath} 
\pgfplotsset{compat=1.18} 
\usepackage{lipsum} 
\usepackage{booktabs} 
\usepackage{siunitx}  
\usepackage{rotating} 
\usepackage{tabularx}
\usepackage{multirow}

\setcopyright{acmlicensed}
\copyrightyear{2026}
\acmYear{2026}
\acmDOI{XXXXXXX.XXXXXXX}
\acmConference[ICSE'26]{IEEE/ACM International Conference on Software Engineering}{April 12-18,
  2026}{Rio de Janeiro}
\acmISBN{978-1-4503-XXXX-X/2018/06}

\begin{document}

\title{A Hybrid, Knowledge-Guided Evolutionary Framework for Personalized Compiler Auto-Tuning}

\author{%
Haolin Pan$^{1,2,3}$,
Hongbin Zhang$^{2}$,
Mingjie Xing$^{2,^{*}}$,
Yanjun Wu$^{2,3}$
}

\affiliation{%
  \vspace{3pt} 
  \textsuperscript{1} Hangzhou Institute for Advanced Study at UCAS, Hangzhou, China\\
  \textsuperscript{2} Institute of Software, Chinese Academy of Sciences, Beijing, China\\
  \textsuperscript{3} University of Chinese Academy of Sciences, Beijing, China
  \city{Beijing}
  \country{China}
}

\email{{panhaolin21@mails.ucas.ac.cn, {hongbin2019,mingjie,yanjun}@iscas.ac.cn}}
\thanks{\textsuperscript{*} Corresponding author.}

\renewcommand{\shortauthors}{Haolin Pan et al.}

\begin{abstract}
Compiler pass auto-tuning is critical for enhancing software performance, yet finding the optimal pass sequence for a specific program is an NP-hard problem. Traditional, general-purpose optimization flags like \texttt{-O3} and \texttt{-Oz} adopt a one-size-fits-all approach, often failing to unlock a program's full performance potential. To address this challenge, we propose a novel \textbf{Hybrid, Knowledge-Guided Evolutionary Framework}. This framework intelligently guides online, personalized optimization using knowledge extracted from a large-scale offline analysis phase. During the offline stage, we construct a comprehensive compilation knowledge base composed of four key components: (1) Pass Behavioral Vectors to quantitatively capture the effectiveness of each optimization; (2) Pass Groups derived from clustering these vectors based on behavior similarity; (3) a Synergy Pass Graph to model beneficial sequential interactions; and (4) a library of Prototype Pass Sequences evolved for distinct program types. In the online stage, a bespoke genetic algorithm leverages this rich knowledge base through specially designed, knowledge-infused genetic operators. These operators transform the search by performing semantically-aware recombination and targeted, restorative mutations. On a suite of seven public datasets, our framework achieves an average of \textbf{11.0\%} additional LLVM IR instruction reduction over the highly-optimized \texttt{opt -Oz} baseline, demonstrating its state-of-the-art capability in discovering personalized, high-performance optimization sequences.
\end{abstract}




\maketitle

\section{Introduction}

\begin{figure}[h]
    \centering
    \includegraphics[width=\columnwidth]{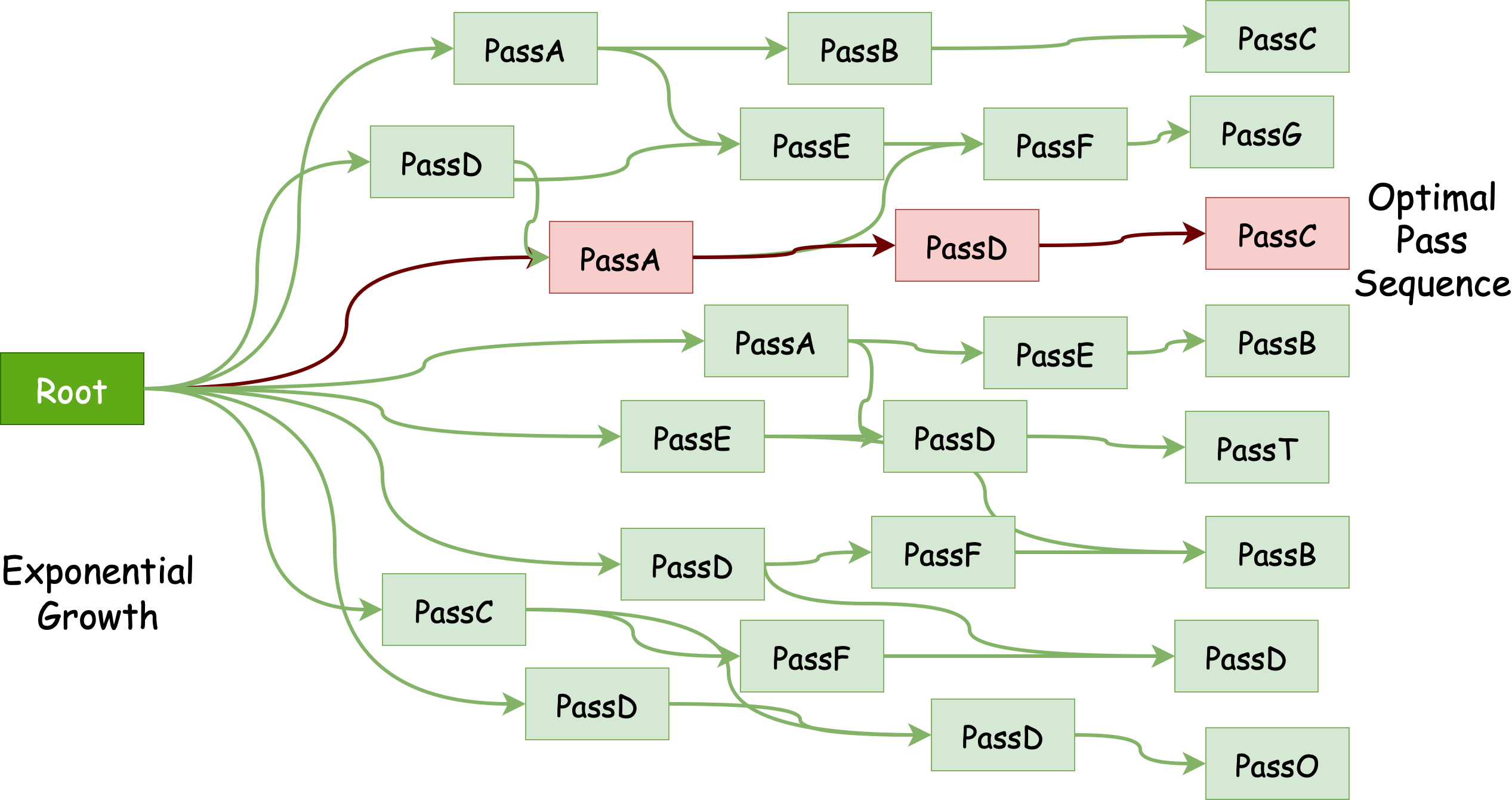} 
    \caption{An illustration of the compiler phase-ordering problem. Starting from the initial program state (Root), the application of different optimization passes (e.g., PassA, PassD) creates a vast, branching search space of possible optimization sequences. The red path highlights optimal pass sequence among countless alternatives.}
    \label{fig:task}
\end{figure}

The relentless pursuit of software performance is a cornerstone of modern computing, fundamentally driven by the compiler's ability to translate high-level source code into efficient machine instructions. Contemporary compilers, such as LLVM, provide a vast arsenal of hundreds of optimization passes. The specific combination and ordering of these passes—commonly known as the "phase-ordering problem"—define a search space that grows exponentially with the number of available passes (see Figure~\ref{fig:task}). Finding the optimal optimization sequence for any given program is a well-known NP-hard problem, posing a significant challenge to consistently achieving peak application performance.

Mainstream compiler optimization strategies face two primary limitations. First, default optimization levels like \texttt{-O3} employ a static, "one-size-fits-all" sequence of passes, failing to adapt to diverse program characteristics and thus leaving significant performance potential untapped \cite{lattner2004llvm}. Second, while auto-tuning methods such as iterative compilation can find superior, program-specific sequences \cite{tpe,cfsat,GA,RIO}, they are often impractical. Their need to explore a vast combinatorial search space through costly trial-and-error compilations results in prohibitive time overheads, making them unsuitable for typical development cycles.

In recent years, machine learning (ML) has emerged as a promising paradigm for addressing the long-standing compiler optimization challenge~\cite{autophase,BOCA,Comptuner,deng2024compilerdream,Reaction_Matching}. ML-based models aim to predict effective optimization strategies directly from program features. While powerful, these approaches often treat the compiler as a black box, thereby overlooking the rich, domain-specific knowledge embedded within its internal design. Critical relationships between optimization passes—such as functional similarity, synergistic collaborations, and antagonistic interferences—are typically ignored. Moreover, these models frequently require extensive, meticulously labeled training data and continue to face persistent challenges in generalizing across new programs, diverse hardware architectures, and evolving compiler versions.

To address these shortcomings, we propose a novel \textbf{Hybrid, Knowledge-Guided Evolutionary Framework}. The core tenet of our methodology is the strategic decoupling of a time-intensive, global knowledge discovery phase from a highly efficient, targeted online optimization phase. In the offline stage, we perform a large-scale, automated analysis of a diverse program corpus. Using a combination of unsupervised learning and graph-based knowledge modeling, we construct a comprehensive and structured compilation knowledge base. This knowledge base encapsulates not only deep insights into program archetypes and pass behaviors but also explicitly models the intricate relationships between them. Subsequently, in the online stage, when a new, unseen program is presented for compilation, our framework leverages this pre-computed knowledge to intelligently steer a bespoke evolutionary algorithm, enabling the rapid and effective discovery of a personalized, high-performance optimization sequence.

We rigorously evaluated our framework on seven widely-used public datasets, using the number of LLVM IR instructions as the metric for code size. The results are compelling: compared to the highly-tuned official LLVM \texttt{opt -Oz} (which aims for minimal code size) optimization level, the personalized sequences generated by our framework achieve an average additional instruction reduction of \textbf{11.0\%}. This significant improvement validates our central hypothesis that intelligently guiding an online search with deeply-mined offline knowledge is an effective path toward solving the compiler optimization problem. The principal contributions of this work are as follows:
\begin{itemize}
    \item We propose a novel, hybrid auto-tuning framework that seamlessly integrates an extensive offline knowledge extraction phase with a rapid online search. This framework automatically constructs a comprehensive compilation knowledge base, which is composed of four key, data-driven components: Pass Behavioral Vectors, behavior Pass Groups, a Synergy Pass Graph, and Prototype Pass Sequences.

    \item We design a set of novel, knowledge-guided genetic operators, including a crossover operator that leverages functional pass clusters for semantic recombination, and a targeted, restorative mutation operator that utilizes the knowledge base for intelligent candidate generation and parallelized online validation.

    \item We conduct extensive experiments, demonstrating that our framework significantly outperforms highly optimized, general-purpose compiler baselines, and consistently achieves state-of-the-art performance in personalized optimization sequence discovery tasks.
\end{itemize}

The remainder of this paper is structured as follows. We begin by surveying related work in Section~\ref{sec:related_work}. Section~\ref{sec:methodology} provides a detailed exposition of our proposed hybrid framework, detailing both the offline knowledge-base construction and the online personalized evolution phases. In Section~\ref{sec:experiments}, we present our comprehensive experimental evaluation, including a comparative performance analysis against state-of-the-art methods and an in-depth ablation study. Finally, we conclude the paper in Section~\ref{sec:conclusion}.

\begin{figure*}[htbp]
    \centering
    \includegraphics[width=\textwidth]{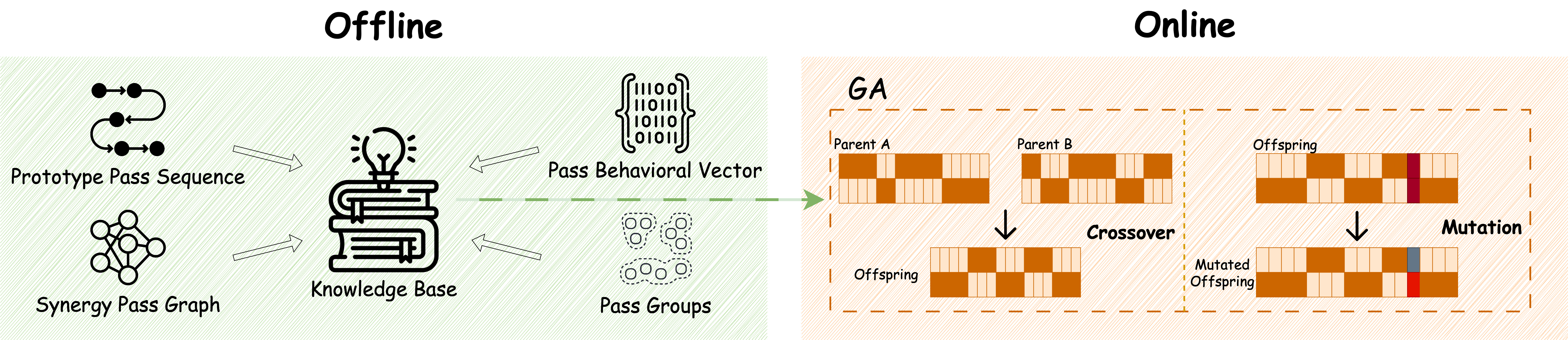} 
    \caption{An overview of the proposed Hybrid, Knowledge-Guided Evolutionary Framework, which is divided into Offline and Online phases.}
    \label{fig:framework_overview}
\end{figure*}

\section{Related Work}
\label{sec:related_work}

The automatic tuning of compiler optimizations, a domain encompassing the phase-ordering and pass selection problems, has been the subject of extensive research for several decades. This section surveys pivotal developments in the field, charting the progression from foundational iterative compilation techniques to machine learning-based predictive models and the nascent application of Large Language Models (LLMs). We conclude by positioning our hybrid framework, highlighting how it synthesizes insights from these preceding paradigms while introducing unique contributions to address their inherent limitations.

\subsection{Iterative Compilation}
\label{ssec:iterative_compilation}

The earliest and most direct methodologies for compiler auto-tuning fall under the umbrella of iterative compilation (IC)~\cite{bodin1998iterative}. The core principle of IC is to treat the compiler as a configurable black box, empirically exploring a vast space of optimization pass sequences for a single program. By repeatedly compiling and evaluating the program with different sequences, an optimal or near-optimal configuration for that specific program-input pair can be discovered. The exploration of this search space has been approached with a diverse array of heuristic search strategies~\cite{cfsat,tpe,RIO,ICMC}, including genetic algorithms~\cite{GA}, particle swarm optimization, and simulated annealing. Frameworks such as OpenTuner~\cite{opentuner} have generalized this approach, providing extensible platforms for autotuning not only compilers but a wide range of configurable software.

While IC is capable of uncovering highly specialized and effective optimization sequences, its primary drawback is the prohibitive time cost associated with its search process. Faced with a combinatorial search space of astronomical size, IC must perform numerous, costly trial-and-error compilations and evaluations for each new program. This high compilation latency renders it impractical for integration into standard software development workflows, where rapid build times are essential. Our work acknowledges the undeniable power of empirical search but fundamentally re-architects its application. Instead of performing this exhaustive exploration at compile-time for every program, we strategically shift the heavy computational burden to a one-time, offline knowledge discovery phase. This phase builds a durable knowledge base from a large program corpus, which then serves to intelligently guide a fast and targeted online search.

\subsection{ML-based Approaches}
\label{ssec:ml_approaches}

To circumvent the significant overhead of iterative compilation, the research community has increasingly turned to ML to create predictive models for compiler optimization~\cite{srtuner, wang2018machine,leather2020machine,BOCA,Comptuner,shahzad2024neural}. The central premise of these approaches is to learn a mapping from a representation of a program's characteristics—its features—to an effective optimization strategy. Once trained on a large dataset of programs and their corresponding effective pass sequences (often derived from an initial IC phase), these models can infer beneficial optimizations for new, unseen programs at a fraction of the cost of a full iterative search.

This paradigm has been explored through various ML techniques. Supervised learning models, for instance, have been trained to predict optimal loop unroll factors or to select from a pre-defined set of promising optimization sequences. A particularly prominent technique in this domain is reinforcement learning (RL)\cite{rotem2021profile}, which frames the phase-ordering challenge as a sequential decision-making problem. In this formulation, an RL agent learns a policy to navigate the optimization space by selecting one pass at a time to apply to the program’s intermediate representation. Seminal works like Autophase\cite{autophase} and the standardized research environment CompilerGym~\cite{CompilerGym} have demonstrated the potential of RL to dynamically construct effective optimization sequences. However, ML-based approaches—especially RL—are often data-hungry and computationally expensive to train. They may also oversimplify the problem by treating optimization passes as independent actions, thus failing to capture the complex and synergistic relationships that exist among them.

\subsection{LLMs in Compilation}
\label{ssec:llm_compilation}

The most recent frontier in automated compiler optimization involves the application of large language models (LLMs)\cite{DeepSeek-R1, jaech2024openai-o1,Search-R1}. Capitalizing on their unprecedented ability to process and generate human-like text, researchers are now exploring the potential of LLMs to "understand" source code and to suggest or generate effective optimization sequences\cite{cummins2024metalargelanguagemodel, italiano2024finding, pan2025compiler,coreset,srtuner}. This emerging area typically involves fine-tuning pre-trained models on vast corpora of code along with associated optimization data. The primary allure of LLMs lies in their potential to bypass traditional, handcrafted feature engineering by learning complex program representations and optimization patterns directly from raw or semi-structured code. Although this field is still in its infancy, it holds strong promise for enabling a more holistic and unified approach to program understanding. Nevertheless, LLM-based methods face several significant challenges, including immense data and computational requirements for training, as well as a general lack of interpretability in their decision-making processes.

\begin{figure*}[htbp]
    \centering
    \includegraphics[width=\textwidth]{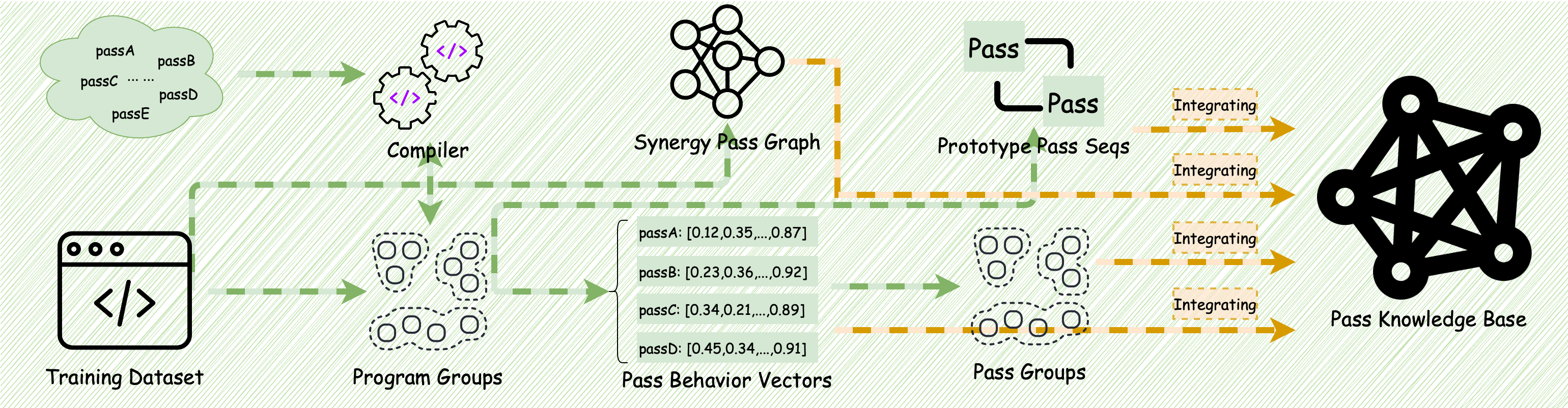} 
    \caption{The offline knowledge extraction pipeline. Starting from a training dataset, we cluster programs to form prototypes. These prototypes are then used to generate a behavioral vector for each compiler pass. The behavioral vectors, in turn, are used to discover functional Pass Groups. In parallel, a Synergy Pass Graph is constructed from empirical data, and Prototype Pass Sequences are evolved for each program prototype. All four of these knowledge components are then integrated into a final, comprehensive Pass Knowledge Base.}
    \label{fig:offline}
\end{figure*}

\subsection{Advantages and Novelty of Our Framework}
\label{ssec:our_framework_advantages}

Our proposed framework is engineered to build upon the foundational insights of these prior works while directly addressing their core limitations through a novel hybrid design. It distinguishes itself by synergistically integrating offline knowledge mining with an efficient, knowledge-guided online evolutionary search.

First, unlike pure iterative compilation, our framework decouples the expensive empirical analysis from the time-critical compilation process of a new program. The offline phase represents a one-time investment that produces a durable and structured \textbf{compilation knowledge base}. This knowledge base does not merely store effective sequences; rather, it encapsulates a deep and generalizable understanding of the optimization space, including quantitative models of pass behavior and their complex inter-relationships. Such a structured representation makes the knowledge more interpretable and broadly applicable than the single-point solutions produced by per-program iterative search approaches.

Second, in contrast to many traditional ML models that learn a direct and often opaque mapping from features to optimizations, our framework constructs an explicit and multi-faceted representation of compiler domain knowledge. By modeling not only what a pass does but also how it relates to other passes, we enable a more nuanced and "semantically-aware" search process. This explicit knowledge representation allows our online algorithm to make more informed decisions, reasoning about the functional roles and collaborative potential of different optimizations.

Finally, our framework's core innovation lies in its \textbf{knowledge-infused genetic operators}. The online evolutionary search is not a blind exploration. It is intelligently initialized with high-quality candidate sequences derived from our offline analysis. Its crossover operator reasons about functional pass clusters, and its targeted mutation operator acts as a "diagnose-and-repair" mechanism. It uses the lightweight behavioral vectors to identify underperforming segments of a sequence and then leverages the knowledge base to intelligently search for superior replacements. This hybrid approach strikes a critical balance, combining the empirical grounding of iterative methods with the efficiency of predictive models, ultimately delivering a solution that is both highly effective and practical for real-world compilation scenarios.

\section{Methodology}
\label{sec:methodology}

The core of our work is a novel hybrid framework designed to automate the discovery of high-performance, personalized compiler optimization sequences. Our methodology is strategically bifurcated into two distinct yet interconnected phases: an extensive, one-time \textbf{Offline Knowledge-Base Construction} phase and a rapid, targeted \textbf{Online Personalized Evolution} phase. This architectural separation is paramount: it allows us to amortize the significant computational cost of deep program analysis and knowledge extraction over a vast corpus of programs, creating a durable and reusable knowledge base. This pre-computed knowledge then serves as a powerful prior to guide the lightweight, efficient online search for an optimal sequence when a new, unseen program is to be compiled.

Figure~\ref{fig:framework_overview} depicts the overall architecture of our framework. The offline phase (Section~\ref{ssec:offline_phase}) processes a large and diverse training set of programs to build a comprehensive compilation knowledge base. This knowledge integrates four key, independently derived components: (1) a quantitative \textbf{Pass behavioral vector} for each optimization; (2) \textbf{Pass Groups} clustering optimizations by functional similarity; (3) a \textbf{Synergy Pass Graph} modeling beneficial sequential interactions; and (4) a library of high-quality \textbf{Prototype Pass Sequences} evolved for distinct program types.

The online phase leverages this rich repository of knowledge. When a new program arrives, a bespoke evolutionary algorithm, equipped with specially designed, knowledge-guided genetic operators, is initiated. Our approach directly uses the offline-generated prototype sequences \textbf{as the initial population} for the evolutionary search, without requiring an initial selection phase. These pre-vetted, high-quality sequences serve as strong starting points. Throughout the search, the algorithm continues querying the knowledge base to perform semantically aware crossover and execute targeted, restorative mutations, enabling rapid convergence on a high-quality, personalized optimization sequence.

\subsection{Offline Knowledge-Base Construction}
\label{ssec:offline_phase}

The offline phase of our framework is designed to systematically distill a large, unstructured corpus of programs, $\mathcal{P}_{train}$, into a structured and actionable compilation knowledge base, $\mathcal{K}$. The core design principle is to pre-compute and store generalizable knowledge, thereby transforming the online optimization task from a blind search into an informed, guided exploration. This knowledge base integrates four key, independently-derived knowledge components. Figure~\ref{fig:offline} provides a high-level overview of this construction pipeline, which consists of four primary stages, each producing a distinct knowledge asset that is subsequently integrated into the final pass knowledge base.

\subsubsection{Pass Behavioral Vector}
\label{sssec:fingerprinting}

To enable effective, data-driven reasoning about compiler passes, we introduce a method to capture their true, context-dependent behavior. The process begins by partitioning the program corpus $\mathcal{P}_{train}$ into $N$ distinct \textbf{program prototypes}. This is achieved by applying K-Means clustering to their Autophase feature vectors~\cite{autophase}, which are normalized prior to clustering to ensure sensitivity to their proportional composition. The resulting clustering model, which we denote as $\mathcal{M}_{prog}$, is preserved for the online phase.

With these prototypes, we define the \textbf{Pass Behavioral Vector} for a pass $\pi$, denoted $\mathbf{f}_\pi$, as an $N$-dimensional vector where each component $(\mathbf{f}_\pi)_i$ quantifies the average effectiveness of $\pi$ on programs of prototype $C_i$. Effectiveness is measured as the percentage reduction in LLVM IR instructions:
\begin{equation}
(\mathbf{f}_\pi)_i = \mathbb{E}_{p \in C_i} \left[ \frac{\text{IR}_{\text{before}}(p) - \text{IR}_{\text{after}}(p, \pi)}{\text{IR}_{\text{before}}(p)} \times 100 \right]
\label{eq:fingerprint}
\end{equation}
where $\text{IR}_{\text{before}}$ and $\text{IR}_{\text{after}}$ are the instruction counts before and after applying pass $\pi$.

\begin{figure*}[ht]
    \centering
    \includegraphics[width=\textwidth]{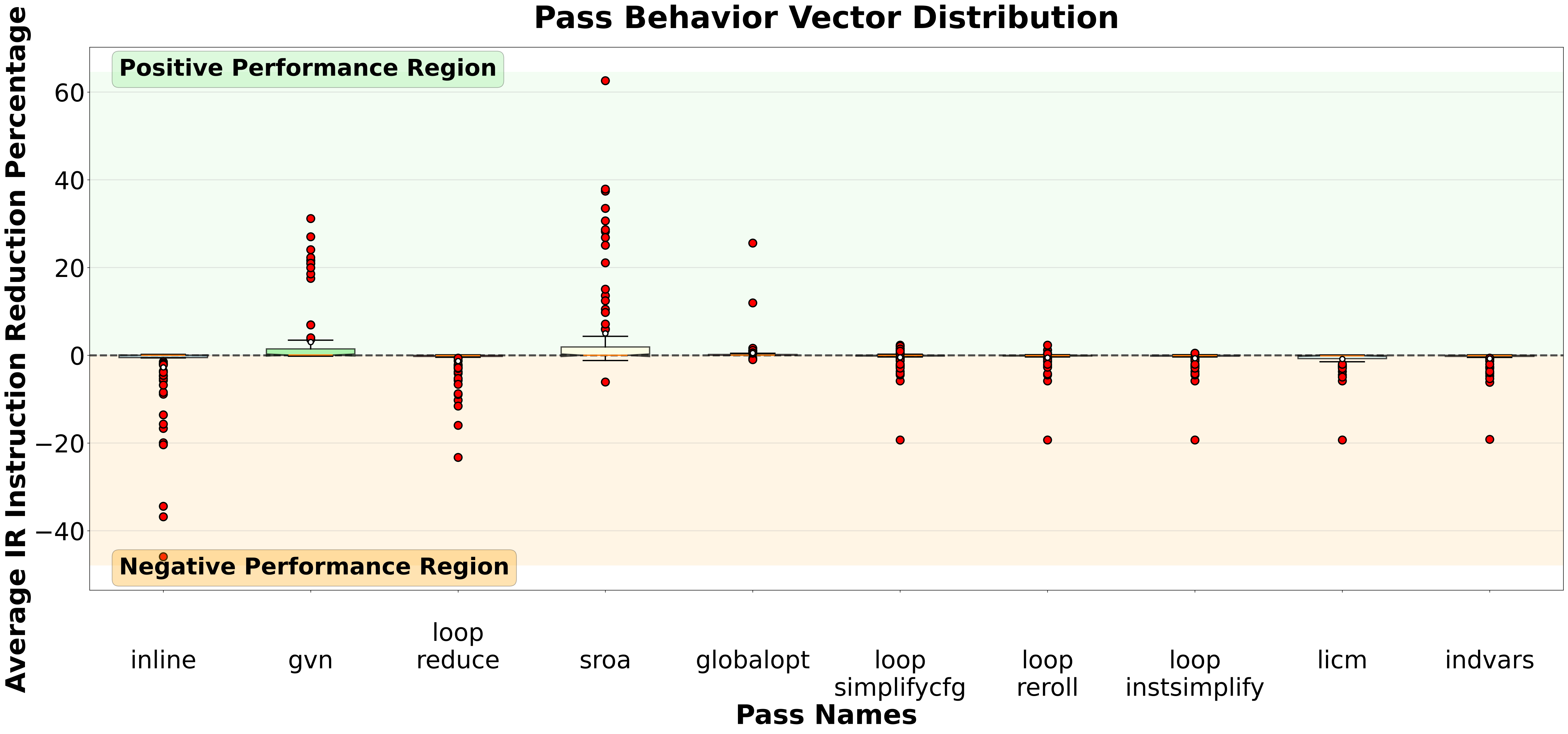} 
    \caption{Distribution of behavioral vector values for a selection of compiler passes. Each boxplot illustrates the performance variation of a single pass across all $N$ program prototypes, highlighting the high context-dependency of pass effectiveness and their diverse behavioral profiles.}
    \label{fig:fingerprint_distribution}
\end{figure*}

The resulting library of behavioral vectors, $\mathcal{F}_{pass}$, provides a rich semantic space for pass analysis. Figure~\ref{fig:fingerprint_distribution} visualizes these vectors for a selection of passes, revealing several critical insights. First, the wide performance distribution of most passes confirms their high \textbf{context-dependency}, validating the need for personalized optimization. Second, passes exhibit distinct \textbf{behavioral profiles}, ranging from the consistently conservative to the high-risk, high-reward. Third, the existence of both positive and negative performance regions for nearly every pass underscores the complexity of the optimization landscape and justifies the necessity of a knowledge-guided search strategy.

Beyond their immediate use in our framework, these vectors serve as a general tool for compiler introspection. They can be used, for instance, to automatically identify risky passes with high performance variance, detect functionally redundant passes via correlation analysis, and provide a quantitative basis for developing more sophisticated, context-aware compiler heuristics.

\subsubsection{Functional Pass Groups}
\label{sssec:pass_groups}

While behavioral vectors provide a detailed view, a higher-level abstraction is needed for efficient reasoning. We aim to identify groups of passes that are functionally similar, as this knowledge can guide substitution and recombination in the online search. We posit that passes exhibiting similar behavioral vectors are likely to perform analogous transformations. Consequently, we discover these \textbf{Pass Groups}, denoted $\mathcal{C}_{pass} = \{G_1, \ldots, G_k\}$, by applying K-Means clustering directly to the set of all pass behavioral vectors, $\mathcal{F}_{pass}$. The algorithm seeks to minimize the total intra-cluster variance:
\begin{equation}
    \underset{\mathcal{C}_{pass}}{\arg\min} \sum_{j=1}^{k} \sum_{\mathbf{f}_\pi \in G_j} ||\mathbf{f}_\pi - \boldsymbol{\mu}_j||^2
    \label{eq:pass_clustering}
\end{equation}
where $\boldsymbol{\mu}_j$ is the centroid of group $G_j$. The optimal number of clusters, $k$, is determined automatically using the Elbow Method. By partitioning the pass space in this manner, we create a set of semantic categories, allowing our online genetic operators to operate on functionally coherent blocks of passes rather than on individual, unrelated ones, leading to a more structured exploration.

\subsubsection{Synergy Pass Graph}
\label{sssec:synergy_graph}

Effective optimization sequences are not merely collections of good passes; their ordering is paramount. To capture the critical, directional relationships between passes, we construct a \textbf{Synergy Pass Graph} \cite{cfsat}. This graph is designed to explicitly model enabling relationships, where one pass creates opportunities for another.

We formally define an ordered pair of passes $(\pi_A, \pi_B)$ as synergistic for a given program $p$ if applying the sequence $\langle \pi_A, \pi_B \rangle$ yields a greater instruction reduction than applying $\pi_B$ alone, under the precondition that $\pi_B$ is itself beneficial. This relationship is captured by the following condition:
\begin{equation}
\text{IR}(p, \langle \pi_A, \pi_B \rangle) < \text{IR}(p, \langle \pi_B \rangle) < \text{IR}(p, \langle \rangle)
\label{eq:synergy}
\end{equation}
where $\text{IR}(p, S)$ denotes the instruction count of program $p$ after applying the optimization sequence $S$, and $\langle \rangle$ represents the empty sequence (the original program).

To build a robust, global model of these interactions, we systematically identify all such synergistic pairs for \textit{every} program $p$ in the entire training corpus $\mathcal{P}_{train}$. These individually discovered pairs are then \textbf{aggregated into a single, comprehensive graph}. The nodes of the Synergy Pass Graph are the passes $\Pi$, and a directed edge exists from $\pi_A$ to $\pi_B$ if the pair was identified as synergistic in at least one training program. The weight of each edge $(\pi_A, \pi_B)$ is proportional to its frequency of occurrence across the corpus, reflecting the strength and generality of the synergistic relationship. The resulting graph, shown as the "Synergy Pass Graph" component in Figure~\ref{fig:offline}, encapsulates a wealth of empirically-validated sequential heuristics, providing a strong, domain-specific prior to guide the construction of logically sound and potent optimization sequences during the online phase.

\subsubsection{Prototype Pass Sequences}
\label{sssec:prototype_sequences}

To mitigate the risk of a cold start and accelerate convergence during the online phase, our framework prepares a library of high-quality initial solutions. This is achieved by evolving \textbf{Prototype Pass Sequences}. The rationale is that a sequence optimized for a class of similar programs serves as a much stronger starting point than a random one. For each of the $N$ program prototypes, we execute a dedicated Genetic Algorithm to find the optimal prototype sequence, $\boldsymbol{\pi}^*_i$. The objective is to maximize the average instruction reduction across all programs in prototype $C_i$:
\begin{equation}
    \boldsymbol{\pi}^*_i = \underset{\boldsymbol{\pi} \in \Pi^L}{\arg\max} \left( \frac{1}{|C_i|} \sum_{p \in C_i} \text{Reduction}(p, \boldsymbol{\pi}) \right)
    \label{eq:prototype_sequence}
\end{equation}
where $\Pi^L$ is the space of all pass sequences of a fixed length $L$, and $\text{Reduction}(p, \boldsymbol{\pi})$ is the performance gain from applying sequence $\boldsymbol{\pi}$ to program $p$. This intensive, prototype-specific offline search yields a library of pre-vetted "seed" sequences, $\mathcal{L}_{proto}$, enabling the online phase to begin its search from regions of the solution space already known to be highly effective.

\subsection{Online Personalized Evolution}
\label{ssec:online_phase}

With the comprehensive, offline-constructed knowledge base $\mathcal{K}$ in place, the framework is ready to perform its primary function: rapidly discovering a high-performance, personalized optimization sequence for a new, unseen program. This is the role of the \textbf{Online Personalized Evolution} phase. The design philosophy of this phase is to be lightweight, fast, and maximally guided by the pre-computed offline knowledge, transforming the search from a blind exploration into an informed, goal-oriented process.

The online process commences with two preparatory steps. First, upon receiving a new program \( p_{new} \), we perform \textbf{rapid profiling} using the pre-trained model \( \mathcal{M}_{prog} \) to instantly classify it into its corresponding program prototype, \( C_{i_{new}} \). This classification, \( i_{new} = \text{Predict}(\mathcal{M}_{prog}, \mathbf{v}_{new}) \), is the critical link that unlocks the relevant segment of our knowledge base. Second, we perform a \textbf{smart initialization} of the evolutionary algorithm's population. Instead of generating random sequences, we leverage the library of empirical prototype sequences, \( \mathcal{L}_{proto} \). We quickly evaluate all sequences in \( \mathcal{L}_{proto} \) on \( p_{new} \) once and select the top-\( k \) best-performing ones to form the initial population, \( P_0 \). This strategy ensures our search begins from a set of high-quality solutions known to be effective for programs of this type.

The core of the online phase is an iterative search performed by our custom-designed Genetic Algorithm (GA). The fitness of any given sequence $\boldsymbol{\pi}$ is formally defined as its percentage reduction in instruction count on $p_{new}$ relative to the \texttt{opt -Oz} baseline:
\begin{equation}
\text{Fitness}(\boldsymbol{\pi}, p_{new}) = \frac{\text{IR}_{\texttt{Oz}}(p_{new}) - \text{IR}_{\boldsymbol{\pi}}(p_{new})}{\text{IR}_{\texttt{Oz}}(p_{new})} \times 100
\label{eq:fitness}
\end{equation}
The defining characteristic of our GA lies in its specialized genetic operators, which we detail below.

\paragraph{Knowledge-Guided Crossover.}
Traditional crossover operators are semantically blind, often disrupting beneficial, co-evolved building blocks (schemata). To overcome this, we introduce a \textbf{Knowledge-Guided Crossover} operator that leverages the functional structure of optimization sequences.

\begin{figure}[t]
    \centering
    \includegraphics[width=\columnwidth]{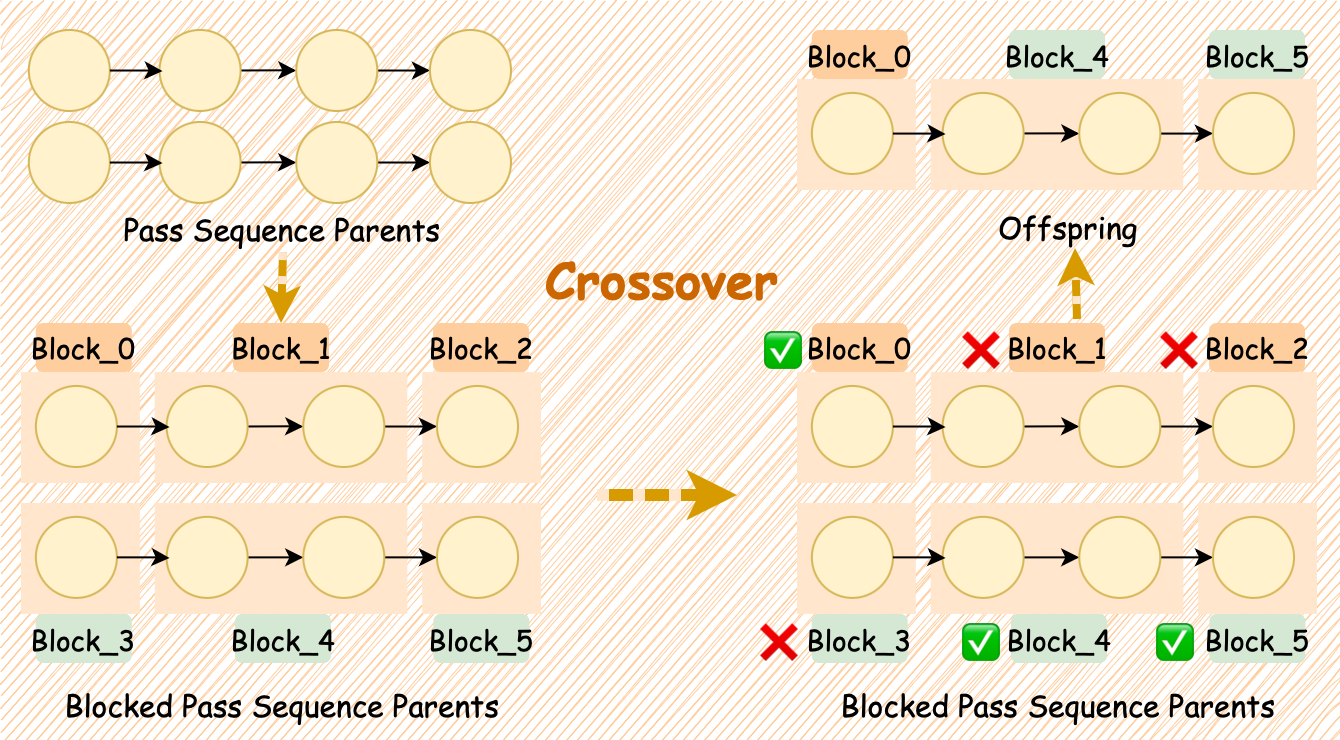} 
    \caption{Conceptual diagram of the Knowledge-Guided Crossover operator. Parent sequences are first segmented into functional blocks. The operator then probabilistically selects and recombines entire blocks, with the selection probability biased by the block's pre-computed effectiveness score (Eq.~\ref{eq:block_score}) for the target program's prototype.}
    \label{fig:crossover}
\end{figure}

As illustrated in Figure~\ref{fig:crossover}, the operator first partitions a parent sequence $\boldsymbol{\pi}$ into a sequence of "functional blocks" $\langle B_1, B_2, \ldots, B_m \rangle$, where each block $B_j$ is a contiguous subsequence of passes belonging to the same functional cluster. The \textit{a priori} effectiveness of a block $B_j$ for the current program's prototype $i_{new}$ is estimated by aggregating the behavioral vector scores of its constituent passes:
\begin{equation}
\text{Score}(B_j, i_{new}) = \sum_{\pi_l \in B_j} (\mathbf{f}_{\pi_l})_{i_{new}}
\label{eq:block_score}
\end{equation}
where $(\mathbf{f}_{\pi_l})_{i_{new}}$ is the $i_{new}$-th component of the vector for pass $\pi_l$. When creating an offspring from two parents, $\boldsymbol{\pi}_a$ and $\boldsymbol{\pi}_b$, for each corresponding block position $j$, the block from parent $\boldsymbol{\pi}_a$ is chosen with a probability proportional to its estimated score. Specifically, the probability of selecting block $B_{j,a}$ from parent $\boldsymbol{\pi}_a$ is given by:
\begin{equation}
P(\text{select } B_{j,a}) = \frac{\text{Score}(B_{j,a}, i_{new})}{\text{Score}(B_{j,a}, i_{new}) + \text{Score}(B_{j,b}, i_{new})}
\label{eq:crossover_prob}
\end{equation}
(Scores are normalized to ensure positivity). This mechanism promotes the preservation and propagation of functionally coherent and historically effective groups of passes, leading to a more meaningful exploration of the search space.

\begin{figure}[t]
    \centering
    \includegraphics[width=\columnwidth]{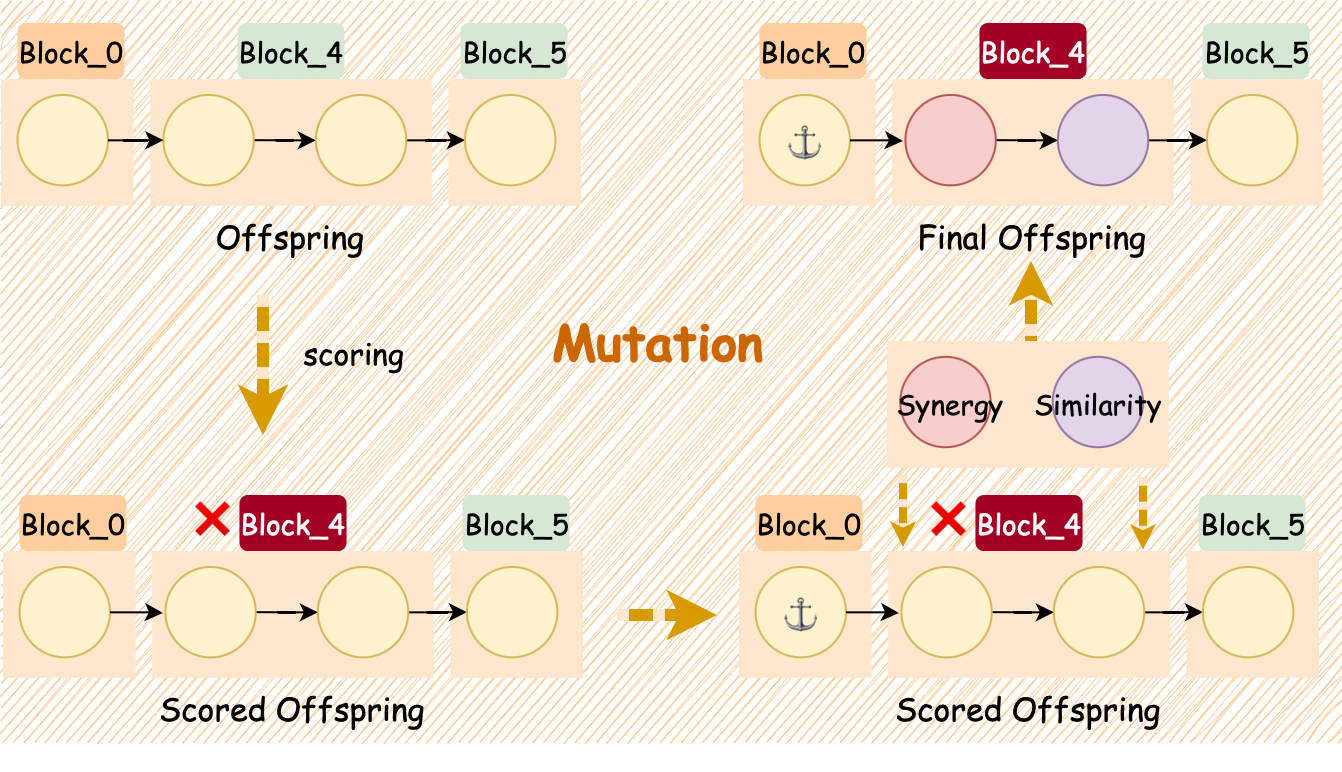} 
    \caption{The Targeted, Restorative Mutation workflow. (1) A sequence is diagnosed using lightweight behavioral vector scores (Eq.~\ref{eq:block_score}) to identify the "weakest link" block. (2) The Pass Knowledge Base is queried to build a high-quality candidate pool based on synergy and similarity. (3) A new, empirically superior block is formed and used to replace the weakest link.}
    \label{fig:mutation}
\end{figure}

\paragraph{Targeted, Restorative Mutation.}
Standard mutation operators introduce random perturbations, a process that is often destructive and inefficient. To address this, we designed a \textbf{Targeted, Restorative Mutation} operator, a sophisticated, multi-stage "diagnose-and-repair" procedure depicted in Figure~\ref{fig:mutation}.

\begin{enumerate}
    \item \textbf{Rapid Diagnosis:} First, the operator diagnoses the sequence to identify its "weakest link." It performs a lightweight scan, calculating the estimated effectiveness score (using Equation~\ref{eq:block_score}) for each functional block in the sequence. The block $B_{worst}$ with the minimum score is identified as the target for mutation:
    \begin{equation}
    B_{worst} = \arg\min_{B_j} \text{Score}(B_j, i_{new})
    \end{equation}

    \item \textbf{Intelligent Candidate Generation:} Second, rather than selecting a random replacement, the operator intelligently constructs a high-quality candidate pool, denoted as $\mathcal{C}_{pool}$. This pool is populated by querying the Pass Knowledge Base and prioritizes passes that exhibit a strong, directed \texttt{synergy} relationship with the pass immediately preceding the target block. To ensure sufficient diversity, the pool is further augmented with passes that share a high \texttt{similarity} to those within the block being replaced (i.e., from the same functional cluster).

    \item \textbf{Empirical Validation and Repair:} Finally, a small set of new candidate blocks, $\{\hat{B}_1, \ldots, \hat{B}_q\}$, are randomly sampled from $\mathcal{C}_{pool}$. These few, promising candidates are then evaluated in parallel via actual compilation on $p_{new}$. The best-performing candidate block, $\hat{B}^*$, is identified:
    \begin{equation}
    \hat{B}^* = \arg\max_{\hat{B}_r} \text{Fitness}(\boldsymbol{\pi}_{\text{new}, r}, p_{new})
    \end{equation}
    where $\boldsymbol{\pi}_{\text{new}, r}$ is the sequence with $B_{worst}$ replaced by $\hat{B}_r$. Only if the fitness of the sequence containing $\hat{B}^*$ is strictly greater than the fitness of the original sequence is the mutation accepted.
\end{enumerate}

This procedure transforms mutation from a blind, random perturbation into a targeted, data-driven, and restorative enhancement, improving search efficiency. After the final generation, the algorithm returns the best sequence found, representing a highly-optimized, personalized solution discovered in a fraction of the time required by traditional iterative methods.

\section{Experiments}
\label{sec:experiments}
\subsection{Main Experiment}
This section presents a comprehensive empirical evaluation of our proposed framework. We begin by detailing the experimental setup, including the environment, datasets, and baseline methods used for comparison. We then present and thoroughly analyze the main experimental results, comparing the performance of our framework against a suite of state-of-the-art, widely recognized compiler auto-tuning techniques.

\subsubsection{Experimental Setup}
\label{ssec:experimental_setup}

\paragraph{Compiler Environment.}All experiments were conducted using the LLVM compiler infrastructure, version 10.0.0. Our framework interacts with LLVM's \texttt{opt} tool to apply sequences of optimization passes and to extract the final instruction counts. The full set of optimization passes available in this LLVM version constitutes the action space for our evolutionary search. The experiments were executed on a high-performance computing system equipped with an AMD EPYC 7763 64-Core Processor, ensuring sufficient computational resources for both the offline knowledge extraction and the parallelized online evaluations.

\paragraph{Datasets.} To ensure a robust and generalizable evaluation, we utilized a comprehensive collection of C and C++ benchmarks, primarily sourced from the CompilerGym environment~\cite{CompilerGym}. The dataset is strategically divided into a large training set for the offline phase and a distinct test set for the final evaluation, as detailed in Table~\ref{tab:dataset_composition}. The training set, composed of \textbf{19,603} program files from sources such as \texttt{github-v0}, \texttt{linux-v0}, and \texttt{poj104-v0}, is exclusively used for building our knowledge base. The test set consists of \textbf{335} unseen program files across seven diverse and curated benchmark suites (\texttt{blas}, \texttt{cbench}, \texttt{chstone}, \texttt{mibench}, \texttt{npb}, \texttt{opencv}, and \texttt{tensorflow}), strictly reserved for evaluating the performance and generalization capabilities of our framework.

\begin{table}[htpb]
\centering
\caption{Dataset Composition detailing the number of programs for training and testing from each source dataset.}
\label{tab:dataset_composition}
\begin{tabularx}{0.48\textwidth}{
        >{\raggedright\arraybackslash}p{0.09\textwidth} |
        >{\centering\arraybackslash}p{0.15\textwidth} |
        >{\centering\arraybackslash}X 
        >{\raggedleft\arraybackslash}X 
        }
\toprule
\textbf{Type} & \textbf{Dataset} & \textbf{Train} & \textbf{Test} \\
\midrule
\multirow{6}{*}{Uncurated} & blas & 133 & 29 \\
 & github-v0  & 7,000 & 0 \\
 & linux-v0  & 4,906 & 0 \\
 & opencv-v0 \cite{opencv} & 149 & 32 \\
 & poj104-v1 \cite{poj} & 7,000 & 0 \\
 & tensorflow-v0 \cite{tensorflow}  & 415 & 90 \\
\midrule
\multirow{4}{*}{Curated} & cbench-v1 \cite{cbench} & 0 & 11 \\
 & mibench-v1 \cite{mibench} & 0 & 40 \\
 & chstone-v0 \cite{chstone} & 0 & 12 \\
 & npb-v0 \cite{npb} & 0 & 121 \\
\midrule
\textbf{Total} & -- & \textbf{19,603} & \textbf{335} \\
\bottomrule
\end{tabularx}
\end{table}

\begin{table*}[htbp] 
\centering
\caption{Average OverOz (\%) Improvement and Time (s) per Program Comparison. Higher OverOz is better. Bold indicates the best performance in each column.}
\label{tab:main_results}
\sisetup{table-align-text-post=false} 
\resizebox{\textwidth}{!}{
\begin{tabular}{@{}l | S[table-format=-2.2] S[table-format=-1.2] S[table-format=1.2] S[table-format=1.2] S[table-format=2.2] S[table-format=1.2] S[table-format=1.2] S[table-format=2.2] r@{}}
\toprule
\textbf{Method} & {\textbf{blas-v0}} & {\textbf{cbench-v1}} & {\textbf{chstone-v0}} & {\textbf{mibench-v1}} & {\textbf{npb-v0}} & {\textbf{opencv-v0}} & {\textbf{tensorflow-v0}} & {\textbf{Avg.}} & {\textbf{Time (s)}} \\ 
\midrule
Opentuner         &  1.60 &  1.99 & 6.46 & 3.33 & 26.19 & 1.76 & 1.29 &  6.09 & 200 \\
GA                & -1.91 &  1.99 & 6.51 & 0.90 & 25.63 & 1.76 & 1.29 &  5.17 & 593 \\
TPE               & -2.24 &  0.97 & 7.60 & 0.20 & 24.62 & 1.46 & 1.23 &  4.83 & 905 \\
RIO               & -2.02 &  0.24 & 4.98 & 3.47 & 23.87 & 0.79 & 1.23 &  4.65 & 200 \\
CompTuner         & -3.06 & -0.65 & 4.38 & -0.45& 22.99 & 0.44 & 1.01 &  3.52 & 10800 \\
BOCA              & -2.36 & -0.16 & 3.18 & -0.69& 22.87 & 1.13 & 1.22 &  3.60 & 3310 \\
\midrule
CFSAT             & 4.60  & 5.80  & 11.90 & 3.70 & 25.90 & \textbf{5.90} & 6.10 & 9.13 & 6 \\
Coreset-NVP       & 2.60  & 3.50  & 9.30  & 1.70 & 9.80  & 5.20 & 6.10 & 5.46 & 11 \\
Compiler-R1       & 5.27  & 5.52  & 9.08  & 6.67 & 22.44  & 4.52 & 5.72 & 8.46 & 29 \\
Autophase (PPO-LSTM) & -1.12 & 5.60 & 4.49 & 4.41 & -4.67 & -0.09 & 0.05 & 1.24 & 3 \\
Autophase (PPO-noLSTM) & -4.77 & -79.69 & -80.90 & -107.33 & -76.69 & -2.32 & -0.76 & -50.35 & \textbf{2} \\
\midrule
\textbf{Ours} & {\textbf{6.15}} & {\textbf{6.80}} & {\textbf{12.57}} & {\textbf{9.02}} & {\textbf{29.74}} & {5.82} & {\textbf{6.93}} & {\textbf{11.00}} & {10} \\
\bottomrule
\end{tabular}%
} 
\end{table*} 

\paragraph{Baselines.}
To rigorously assess the effectiveness of our framework, we compare it against a wide array of established baseline methods. These baselines represent the state-of-the-art in both search-based iterative compilation and machine learning-based approaches. The selected baselines include:
    \begin{itemize}
        \item \textbf{Iterative and Search-Based Methods:} OpenTuner~\cite{opentuner}, a standard Genetic Algorithm (GA)~\cite{GA}, TPE~\cite{tpe}, RIO~\cite{RIO}, CompTuner~\cite{Comptuner}, and BOCA~\cite{BOCA}. These methods represent various heuristic search strategies applied to the phase-ordering problem.
        
        \item \textbf{Machine Learning-Based Methods:} CFSAT~\cite{cfsat} and Coreset-NVP~\cite{coreset}, which are recent, strong performers in this domain, as well as Autophase~\cite{autophase} with its reinforcement learning PPO-LSTM and PPO-noLSTM variants.

        \item \textbf{Large Language Model-Based Method:} To position our work in relation to the latest advancements, we include \textbf{Compiler-R1}~\cite{pan2025compiler} as a representative LLM-based baseline. This method leverages a large language model to automatically generate optimization sequences. In our experiments, the timing results for Compiler-R1 were measured on a system equipped with four NVIDIA H100 GPUs, capturing the total time required for model inference using Qwen-7B, along with the subsequent compiler tool invocation.
\end{itemize}

The primary baseline for performance comparison is LLVM's official size-optimization level, \texttt{opt -Oz}.

\paragraph{Evaluation Metric.}
The primary performance metric used in our evaluation is the \textbf{Average OverOz (\%) Improvement}. This metric quantifies the additional percentage reduction in the number of LLVM IR instructions achieved by a given method compared to the \texttt{opt -Oz} baseline. It is formally defined as:
$$
\text{OverOz (\%)} = \frac{1}{|\mathcal{P}_{test}|} \sum_{p \in \mathcal{P}_{test}} \frac{I_{\texttt{Oz}}(p) - I_{\boldsymbol{\pi}^*}(p)}{I_{\texttt{Oz}}(p)} \times 100
$$
where $\mathcal{P}_{test}$ is the set of programs in a test dataset, $I_{\texttt{Oz}}(p)$ is the instruction count of program $p$ after applying the \texttt{opt -Oz} sequence, and $I_{\boldsymbol{\pi}^*}(p)$ is the instruction count after applying the optimization sequence $\boldsymbol{\pi}^*$ found by the method being evaluated. A higher OverOz value indicates a better optimization performance.

\begin{table*}[ht]
\centering
\caption{Ablation study results showing the average percentage of IR instruction reduction over \texttt{opt -Oz}. Each row represents a variant of our method, and columns correspond to different benchmark suites. The final column shows the average performance across all datasets. Higher values are better.}
\label{tab:ablation_results}
\resizebox{\textwidth}{!}{%
\begin{tabular}{l|ccccccc|c}
\toprule
\textbf{Method Variant} & \textbf{cbench-v1} & \textbf{chstone-v0} & \textbf{mibench-v1} & \textbf{npb-v0} & \textbf{blas-v0} & \textbf{opencv-v0} & \textbf{tensorflow-v0} & \textbf{Average} \\
\midrule
\textbf{Full (Our Method)} & \textbf{6.80} & \textbf{12.57} & \textbf{9.02} & \textbf{29.74} & \textbf{6.15} & \textbf{5.82} & \textbf{6.93} & \textbf{11.0} \\
\midrule
Random Init & 4.93 & 8.99 & 6.36 & 27.94 & 2.21 & 4.92 & 6.45 & 8.8 \\
No-KC (No Knowledge Crossover) & 6.60 & 11.69 & 7.90 & 29.51 & 5.62 & 5.68 & 6.75 & 10.5 \\
No-KM (No Knowledge Mutation) & 6.42 & 10.83 & 7.66 & 29.19 & 5.18 & 5.68 & 6.75 & 10.2 \\
No Knowledge & -4.13 & 0.19 & -0.95 & 16.81 & -1.32 & 1.97 & 2.22 & 2.1 \\
\bottomrule
\end{tabular}%
}
\end{table*}

\subsection{Results}
\label{ssec:main_results}

Table~\ref{tab:main_results} presents the main experimental results, comparing the performance of our framework against all baseline methods across the seven test datasets. The table reports the Average OverOz (\%) improvement for each method on each dataset, the overall average improvement, and the average time taken per program to find the optimization sequence.

\paragraph{Overall Performance.}
The results clearly demonstrate the superior performance of our proposed framework. As shown in the "Avg." column, our method achieves an average OverOz improvement of \textbf{11.00\%}, significantly outperforming all other baselines. This represents a substantial advancement over the next-best method, CFSAT, which achieved a 9.13\% average improvement. A crucial factor contributing to this performance gap is the comprehensive nature of the search space explored. Many search-based baselines, such as TPE, BOCA, CompTuner, and Opentuner, primarily focus on selecting an effective \textit{combination} (or subset) of passes, often applying them in a fixed or heuristically determined order. In contrast, our evolutionary framework is explicitly designed to simultaneously optimize both the selection of passes and their precise sequential ordering. This holistic approach allows our method to uncover complex, synergistic interactions that are inaccessible to methods that decouple these two tightly intertwined aspects of the phase-ordering problem. This strong overall performance underscores the effectiveness of our hybrid, knowledge-guided approach in consistently finding high-quality optimization sequences.

\paragraph{Performance on Individual Datasets.}
Our framework's strength is not limited to its average performance; it consistently delivers top-tier results on individual benchmark suites. It achieves the best performance on five out of the seven datasets: \texttt{blas} (6.15\%), \texttt{cbench} (6.80\%), \texttt{mibench} (9.02\%), \texttt{npb} (29.74\%), and \texttt{tensorflow} (6.93\%). The particularly strong results on computationally intensive benchmarks like \texttt{npb} further underscore the advantage of our holistic search, as such programs are often highly sensitive to the precise ordering of optimizations. In contrast, many traditional search-based methods and even some ML-based approaches show negative results on certain datasets, indicating that their found sequences were worse than the default \texttt{opt -Oz}, a pitfall our robust framework successfully avoids.

\paragraph{Efficiency Analysis.}
In addition to its superior optimization quality, our framework demonstrates a remarkable balance of efficiency and effectiveness. The reported timings in Table~\-ref{tab:main\_results} represent the average time required for each method to converge on its final reported solution. Our framework requires an average of only \textbf{10 seconds} per program to find its high-quality sequences.

This efficiency is particularly notable when contextualized within different methodological categories. When compared to other \textbf{iterative search-based methods}, our framework is not only the most effective but also among the fastest. Traditional iterative compilers like CompTuner (10800s) and BOCA (3310s) require orders of magnitude more time to converge, often because they intertwine their search process with on-the-fly model training or complex heuristic evaluations. Our framework, by contrast, front-loads all expensive learning into the one-time offline phase, making the online search a pure, guided evaluation process. When compared to \textbf{ML-based predictive methods}, such as CFSAT (6s), our framework's slightly longer search time is an expected trade-off for its higher optimization quality. Methods like CFSAT primarily rely on a single, fast forward pass through a pre-trained model to predict a sequence. Our approach, while also guided by a pre-trained knowledge base, is fundamentally an \textbf{iterative compilation process} that performs multiple, empirical evaluations on the target program. This iterative refinement allows our framework to fine-tune a sequence specifically for the given program, uncovering optimizations that a pure predictive model might miss. Consequently, our framework achieves the best performance among all iterative methods while maintaining a convergence time that is competitive and practical, positioning it as a highly viable solution for real-world compiler optimization challenges.

\subsection{Ablation Study}
\label{sec:ablation_study}

To rigorously evaluate the efficacy of the individual components within our knowledge-guided evolutionary framework, we conducted a comprehensive ablation study. The primary objective of this study is to dissect the framework and quantify the performance contribution of each of our proposed knowledge-infused mechanisms: (1) the smart, prototype-based initialization, (2) the knowledge-guided crossover operator, and (3) the targeted, restorative mutation operator. By systematically disabling these components, we can isolate their impact and validate their necessity.

\subsubsection{Experimental Setup}

To systematically evaluate the contributions of our framework's components, we conduct a comprehensive ablation study. Our complete proposed method, denoted as \textbf{Full}, is compared against four specialized, ablated variants. First, to measure the direct benefit of a strong starting point, the \textbf{Random-Init} variant disables our smart initialization strategy, replacing the high-quality empirical prototype sequences with a randomly generated population while all other knowledge-guided operators remain active. Second, to isolate the contribution of our semantically-aware block recombination, the \textbf{No-KC} (No Knowledge Crossover) variant substitutes our knowledge-guided crossover with a standard, semantically-blind single-point crossover, while retaining smart initialization and knowledge-guided mutation. Similarly, the \textbf{No-KM} (No Knowledge Mutation) variant measures the impact of our "diagnose-and-repair" mechanism by replacing our targeted, restorative mutation with a conventional random mutation operator. Finally, the \textbf{No-Knowledge} variant serves as our primary baseline, disabling all knowledge-guided components, thereby allowing us to quantify the total performance gain attributable to our entire knowledge-guidance paradigm. For all experiments, core GA hyperparameters were held constant for a fair comparison, and performance was measured as the average percentage reduction in LLVM IR instructions relative to the highly-optimized \texttt{-Oz} baseline.

\subsubsection{Results and Analysis}

\begin{figure*}[t]
    \centering
    \includegraphics[width=\textwidth]{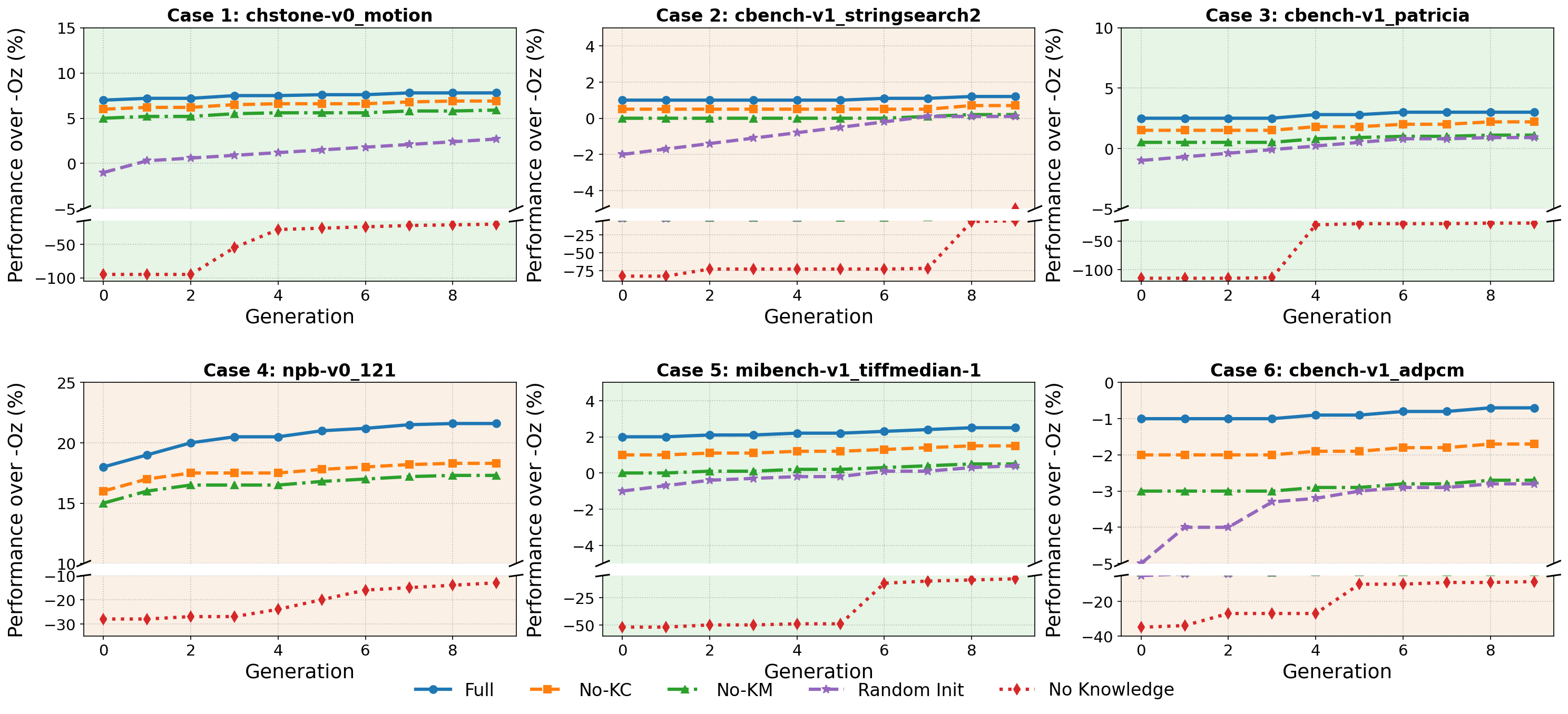} 
    \caption{Convergence curves on representative benchmarks, demonstrating the critical role of our knowledge-guided components. Our \textbf{Full} method shows rapid convergence to superior solutions, while disabling knowledge—particularly smart initialization—results in a significantly less efficient search.}
\label{fig:convergence_curves}
\end{figure*}

The results of our ablation study are summarized in Table~\ref{tab:ablation_results}, which reports the final converged performance of each variant. To provide deeper insight into the search process itself, Figure~\ref{fig:convergence_curves} illustrates the convergence behavior of each method on a selection of representative benchmarks. 

The convergence curves clearly visualize the impact of our knowledge-guided components. The \textbf{Full} method consistently starts from a high initial performance and converges rapidly to the best solution, a direct result of its smart initialization and efficient search operators. The ablated variants exhibit progressively degraded convergence behavior, with the \texttt{No Knowledge} often starting from a significantly negative performance and struggling to improve. This visual evidence reinforces the quantitative results in the table, demonstrating not only that our full method achieves superior final performance, but also that its search process is fundamentally more efficient and better-directed. The subsequent analysis will focus on the final performance metrics from the table.

\textbf{The Impact of Smart Initialization.} Comparing the \textbf{Full} method with the \textbf{Random Init} variant reveals the substantial benefit of our smart initialization strategy. Across all benchmarks, replacing prototype-based initialization with random initialization leads to a significant performance drop, with the average reduction declining from approximately 11.0\% to 8.8\% (a relative decrease of 20\%). For instance, on `chstone-v0`, performance falls from 12.41\% to 8.99\%. This confirms that starting the search from a set of high-quality, pre-vetted sequences provides a critical advantage, allowing the GA to focus its efforts on refining already strong solutions rather than discovering them from scratch.

\textbf{The Contribution of Knowledge-Guided Operators.} The contributions of our specialized crossover and mutation operators become clear when comparing the \textbf{Full} method to the \textbf{No-KC} and \textbf{No-KM} variants. Disabling the knowledge-guided crossover (\textbf{No-KC}) results in a consistent, though moderate, performance degradation. For example, on \texttt{mibench-v1}, performance drops from 9.02\% to 7.90\%. This indicates that our semantically aware block recombination effectively preserves and propagates beneficial building blocks, contributing to a more efficient search. The removal of knowledge-guided mutation (\textbf{No-KM}) causes an even more pronounced decline than removing the intelligent crossover. On \texttt{chstone-v0}, performance decreases from 12.57\% to 10.83\%. This suggests that our ``diagnose-and-repair'' mutation mechanism, which intelligently targets and fixes weak points in a sequence, is a more critical contributor to finding high-quality solutions than the crossover operator. It actively enhances solutions, whereas crossover primarily recombines existing genetic material.

\textbf{The Overall Value of Knowledge Guidance.} The most dramatic performance drop occurs in the \textbf{Standard GA} variant, where all knowledge-guided components are disabled. In this scenario, the performance plummets, even becoming significantly worse than the \texttt{opt -Oz} baseline on several benchmarks (e.g., -4.13\% on `cbench-v1`). This stark contrast highlights the immense value of our complete knowledge-guidance paradigm. Without the priors from the offline knowledge base, a standard GA struggles to navigate the vast and complex optimization space effectively within a limited time budget. The consistently positive and superior results of our \textbf{Full} method underscore the conclusion that a holistic, knowledge-driven approach is essential for achieving state-of-the-art performance in personalized compiler optimization.


\section{Conclusion}
\label{sec:conclusion}

In this paper, we presented a \textbf{Hybrid, Knowledge-Guided Evolutionary Framework} for the compiler phase-ordering problem, designed to balance optimization quality with practical compilation overhead. Our approach is founded on the strategic decoupling of an extensive offline knowledge extraction phase from a knowledge-driven online search. The offline phase automatically constructs a multi-faceted Pass Knowledge Base from a large program corpus. This knowledge base provides a quantitative, data-driven understanding of the optimization space by modeling: (1) the context-dependent performance of passes via Behavioral Vectors, (2) their functional relationships through Pass Groups, (3) their sequential dependencies in a Synergy Graph, and (4) effective initial solutions via Prototype Sequences. This pre-computed knowledge is then utilized by a bespoke online evolutionary algorithm, whose knowledge-infused genetic operators enable an efficient and structured exploration of the solution space.

Empirically validated on seven public datasets, our framework achieved an average additional instruction count reduction of 11.0\% over the highly-optimized \texttt{-Oz} baseline, confirming its practicality and effectiveness. Future work will extend this framework to multi-objective optimization (e.g., execution time, energy) and explore integrating learned program representations from large language models. In conclusion, our work demonstrates that a deep, offline analysis of compiler behavior, when systematically leveraged by a lightweight online search, provides a promising path towards more intelligent and adaptive self-tuning compilers.

\bibliographystyle{ACM-Reference-Format}
\bibliography{ref}

\end{document}